# Title: The Relationship between Major Depression Symptom Severity and Sleep Collected Using a Wristband Wearable Device: Multi-centre Longitudinal Observational Study

Yuezhou Zhang[1], MSc; Amos A Folarin[1,2,8], PhD; Shaoxiong Sun[1], PhD; Nicholas Cummins[1], PhD; Rebecca Bendayan[1,9], PhD; Yatharth Ranjan[1], MSc; Zulqarnain Rashid[1], PhD; Pauline Conde[1], BSc; Callum Stewart[1], MSc; Petroula Laiou[1], PhD; Faith Matcham[3], PhD; Katie White[3], BSc; Femke Lamers[4], PhD; Sara Siddi[5], PhD; Sara Simblett[3,8], PhD; Inez Myin-Germeys[6], PhD; Aki Rintala[6], MSc; Til Wykes[3,8], PhD; Josep Maria Haro[5], MD; Brenda WJH Pennix[4], PhD; Vaibhav A Narayan[7], PhD; Matthew Hotopf[3,8], PhD; Richard JB Dobson[1,2,9], PhD; RADAR-CNS consortium[10]

[1]The Department of Biostatistics and Health informatics, Institute of Psychiatry, Psychology and Neuroscience, King's College London, London, UK
[2]Institute of Health Informatics, University College London, London, UK
[3]Institute of Psychiatry, Psychology and Neuroscience, King's College London, London, UK
[4]Department of Psychiatry, Amsterdam UMC, Vrije Universiteit and GGZinGeest, Amsterdam, The Netherlands
[5]Parc Sanitari Sant Joan de Déu, CIBERSAM, Universitat de Barcelona, Sant Boi de Llobregat, Barcelona, Spain
[6]Centre for Contextual Psychiatry, Department of Neurosciences, KU Leuven, Leuven, Belgium
[7]Janssen Research and Development LLC, Titusville, NJ, USA
[8]South London and Maudsley NHS Foundation Trust, London, UK
[9]NIHR Biomedical Research Centre at South London and Maudsley NHS Foundation Trust and King's College London, London, UK
[10]www.radar-cns.org

**Corresponding author**
Richard Dobson
Department of Biostatistics & Health Informatics
SGDP Centre, IoPPN
King's College London
Box PO 80
De Crespigny Park, Denmark Hill
London
SE5 8AF
Email: richard.j.dobson@kcl.ac.uk
Telephone: +44(0) 20 7848 0473

## Abstract

**Background:** Sleep problems tend to vary accordingly to the course of the disorder in individuals with mental health problems. Research in mental health has implicated sleep pathologies with depression. However, the gold standard for sleep assessment, polysomnography (PSG), is not suitable for long-term, continuous, monitoring of daily sleep, and methods such as sleep diaries rely on subjective recall, which is qualitative and inaccurate. Wearable devices, on the other hand, provide a low-cost and convenient means to monitor sleep in home settings.

**Objective:** The main aim of this study was to devise and extract sleep features, from data collected using a wearable device, and analyse their correlation with depressive symptom severity and sleep quality, as measured by the self-assessed Patient Health Questionnaire 8-item (PHQ-8).

**Methods:** Daily sleep data were collected passively by Fitbit wristband devices, and depressive symptom severity was self-reported every two weeks by the PHQ-8. The data used in this paper included 2,812 PHQ-8 records from 368 participants recruited from three study sites in the Netherlands, Spain, and the UK. We extracted 21 sleep features from Fitbit data which describe the participant's sleep in the following five aspects: sleep architecture, sleep stability, sleep quality, insomnia, and hypersomnia. Linear mixed regression models were used to explore associations between sleep features and depressive symptom severity. The z-test was used to evaluate the significance of the coefficient of each feature.

**Results:** We tested our models on the entire dataset and individually on the data of three different study sites. We identified 16 sleep features that were significantly ($P < .05$) correlated with the PHQ-8 score on the entire dataset, among them, awake proportion ($z = 5.45$, $P < .001$), awakening times ($z = 5.53$, $P < .001$), insomnia ($z = 4.55$, $P < .001$), mean sleep offset time ($z = 6.19$, $P < .001$) and hypersomnia ($z = 5.30$, $P < .001$) were the top 5 features ranked by z-test statistics. Associations between sleep features and the PHQ-8 score varied across different sites, possibly due to the difference in the populations. We observed that many of our findings were consistent with previous studies, which used other measurements to assess sleep, such as PSG and sleep questionnaires.

**Conclusion:** Although consumer wearable devices may not be a substitute for PSG to assess sleep quality accurately, we demonstrate that some derived sleep features extracted from these wearable devices show potential for remote measurement of sleep as a biomarker of depression in real-world settings. These findings may provide the basis for the development of clinical tools to passively monitor disease state and trajectory, with minimal burden on the participant.

## Introduction

According to the report of the World Health Organization (WHO), the total number of people with depression was estimated to exceed 300 million in 2015, equivalent to 4.4% of the world's population [1]. There are several depression-related adverse outcomes, including premature mortality [2], declining in quality-of-life [3], and loss of occupational function [4]. Depression is also a major contributor to deaths by suicide. Close to 800 000 people die due to suicide every year [1].

Sleep disturbances are prevalent among depression patients; more than 90% of patients with depression reported poor sleep quality [5]. Sleep disturbances cover a wide range of different symptoms and disorders including, insomnia, hypersomnia, excessive daytime sleepiness, circadian rhythm disturbance [6]. Insomnia and sleep quality have been observed to be bi-directionally related to depression in several longitudinal studies [6]. Hypersomnia is more frequently present in depressive episodes of bipolar patients [7, 8]. Changes in sleep architecture, such as reduced deep sleep, increased REM sleep and shortened REM latency, are also significant predictors of depression [9, 10].

The gold standard for sleep evaluation is polysomnography (PSG), which involves several physiological measurements, electroencephalogram (EEG), electrocardiogram (ECG), electromyogram (EMG), and accelerometers (ACC) [11]. Using PSG to assess sleep is time-consuming, expensive, and labour-intensive, requiring dedicated equipment and separate laboratory rooms as well as experts to analyse the physiological signals. Since depression can affect patients for an extended period, long-term monitoring of sleep quality is essential. Due to the above shortcomings, PSG is not suitable for long-term sleep monitoring [12]. Sleep questionnaires, such as the Pittsburgh Sleep Quality Index (PSQI) [13], are another useful method to assess sleep. This method relies on the self-reporting of subjective factors, like low recall of sleep, that may affect the accuracy of the assessment [14].

Several recent studies have used wearable devices to estimate sleep quality and sleep-related parameters [15-18] and analysed the relationship between sleep and depression [19-21]. Miwa et al. estimated sleep quality by detecting roll-over movements during sleep and observed that there was a significant difference in sleep quality between non-depressed and depressed people [19]. Mark et al. estimated the sleep duration of 40 information workers for 12 days using a Fitbit wristband and found that sleep duration was positively correlated with mood [20]. DeMasi et al. found that sleep was significantly related to changes in depressive symptoms [21]. These studies have mostly been performed on the single-centre and relatively small datasets (number of participants less than 100). Moreover, most of these studies only used some basic sleep parameters, such as sleep duration; detailed information on sleep architecture, sleep patterns, and stability of sleep was not considered. The relationship between detailed sleep features, as estimated from data supplied by wearable devices, and depression is yet to be fully explored.

The first aim of this study was to design more sleep-related features, from wearable device data, that reflect the sleep architecture, sleep stability, sleep quality, and sleep disturbances (insomnia and hypersomnia) of a participant. The second aim was to perform analysis on a relatively large and multi-site dataset to explore associations between these sleep features and depressive symptom severity. The third aim was to compare our findings with previous studies which used other measurements to assess sleep, such as PSG and sleep questionnaires.

## Methods

### Dataset

The data we used in this paper were collected from a major EU Innovative Medicines Initiative (IMI) research project, Remote Assessment of Disease and Relapse - Central Nervous System (RADAR-CNS) [22]. This project aims to investigate the use of remote measurement technologies to monitor people with depression, epilepsy, and multiple sclerosis in real-world settings. The study protocol for the depression component (Remote Assessment of Disease and Relapse – Major Depressive Disorder; RADAR-MDD) is described in detail in [23]. The RADAR-MDD project aims to recruit 600 participants with a recent history of depression in three study sites (King's College London (KCL; UK), Vrije Universiteit Medisch Centrum (VUmc; Amsterdam, The Netherlands), and Centro de Investigación Biomédican en Red (CIBER; Barcelona, Spain). Recruitment procedures vary slightly across sites, eligible participants are identified either through existing research cohorts (in KCL and VUmc) who had given consent to be contacted for research purposes or through mental health services (in KCL and CIBER) [23]. Participants from KCL and VUmc are community-based, while the participants from CIBER come from a clinical population. As part of the study, participants are asked to install several remote monitoring technology (RMT) apps and use an activity tracker for up to 2 years of follow-up. Many categories of passive and active data are being collected and uploaded to an open-source platform, RADAR-base [24]. In this paper, we focus on the sleep and Patient Health Questionnaire 8-item (PHQ-8) data [25].

### *Sleep data*

According to the American Academy of Sleep Medicine (AASM) manual for the scoring of sleep and associated events, sleep can be divided into two phases: rapid eye movement (REM) sleep and non-REM (NREM) sleep; and NREM sleep can be subdivided into N1, N2, N3 stages according to characteristic patterns of brain waves collected by PSG [11]. In our project, the daily sleep records of participants were collected by the Fitbit Charge 2 or Charge 3. An entire night's sleep is divided into four stages: "Awake", "Light", "Deep", and "REM". The "Light" stage provides estimates for the N1 and N2 stages in PSG, while the "Deep" stage estimates for the N3 stage in PSG. According to several validation studies of Fitbit, the Fitbit wristband had limited specificity in sleep stages estimation [26-28]. Therefore, in this study, we were not expecting the Fitbit devices to provide as accurate information as PSG would have provided. However, the Fitbit devices were deemed sensitive enough to detect changes in sleep-wake states [26-28]; therefore, the provided sleep stage information could be utilised to determine estimates for detailed sleep parameters based on known sleep pathology.

### *PHQ-8 data*

The variability of each participant's depressive symptom severity was measured via the PHQ-8, conducted by mobile phone every two weeks. The questionnaire contains eight questions, with the score of each sub-item ranging from 0 to 3. The total score (range 0 - 24) of all sub-items is the PHQ-8 score which can evaluate depressive symptom severity of the participant for the past two weeks. A PHQ-8 score ≥10 is the most commonly recommended cut-point for "clinically significant" depressive symptoms [25], i.e., if the PHQ-8 of a participant is ≥10, the participant is likely to have had depressive symptoms in the previous two weeks. In the PHQ-8, sub-item 3 is specifically corresponding to sleep. The content of sub-item 3 is: "Trouble falling or staying asleep, or sleeping too much" [25]. A higher score in sub-item 3 indicates worse self-reported sleep in the past two weeks. For reading convenience, we denoted the score of sub-item 3 as the sleep subscore in this paper.

### *Socio-demographics*

Socio-demographic of participants were collected during the enrolment session. According to previous studies on the associations between depression and socio-demographic characteristics [29,30], we considered baseline age, gender, education level and annual income as potential confounding variables in our analyses. Due to the different educational systems in different countries, we simply divided the education level into 2 levels: degree (or above) and below degree. The annual income levels of Spain and the Netherlands had been transformed into equivalent British levels.

## Feature extraction

### Feature window size

For each PHQ-8 record, we extracted sleep features from a time window prior to the PHQ-8 completion time. Since the PHQ-8 score is used to represent the depressive symptom severity of the participant for the past two weeks, we set the feature window size to 2 weeks which was denoted as $\Delta t$ in this paper.

### Sleep Features

According to known sleep pathology and our experience, 21 sleep features extracted in this paper were divided into the following five categories (Table 1): *sleep architecture*, representing the basic and cyclical patterns of sleep; *sleep stability*, representing the variance of sleep in the feature window; *sleep quality*, measures relating to total sleep and wake times; *insomnia*, trouble falling or staying asleep; *and hypersomnia*, excessive sleepiness.

#### Sleep Architecture

The features of sleep architecture were intended to describe the basic and cyclical patterns of sleep. Therefore, we extracted some features similar to those in the PSG report (sleep duration, time in bed, and REM latency) [31], and features of the proportions of all sleep stages.

*Sleep duration*. The sleep duration of one night is defined as the sum of all "Non-Awake" stages ("Light", "Deep", and "REM") [31]. The mean sleep duration in $\Delta t$ was denoted as *Sleep_dur*.

*Time in bed*. Time in bed of one night is defined as the sum of all sleep stages ("Awake", "Light", "Deep", and "REM") of the entire night [31]. The mean sleep duration in $\Delta t$ was denoted as *Time_bed*.

*The proportion of each sleep stage*. The proportion of each sleep stage is defined as the ratio of the time in this sleep stage to the time in bed of the entire night. Different sleep stages have different functions and can reflect the quality of sleep. Deep sleep is considered essential for memory consolidation [32], and REM sleep favours the preservation of memory [33]. A previous sleep report has shown that more deep sleep and fewer awakenings represent better sleep quality [31]. Therefore, we extracted the mean proportion of these four sleep stages in $\Delta t$, and denoted them as *Deep_pct*, *Light_pct*, *REM_pct*, *Awake_pct*, respectively. Besides, the combination of deep and light sleep is NREM sleep. The mental activity that occurs in NREM and REM sleep is a result of two different mind generators, which also explains the difference in mental activity [34]. So, we extracted the mean proportion of NREM sleep in $\Delta t$ which was denoted as *NREM_pct*.

*REM latency*. Previous literature has shown that shortened REM latency can be considered as a biological mark of depression relapse [9]. REM latency is defined as the interval between sleep onset and the occurrence of the first REM stage. The mean REM latency in $\Delta t$ was denoted as *REM_L*.

### Sleep Stability

The features in this category were used to estimate the variance of sleep during $\Delta t$. We extracted the standard deviation of sleep duration, sleep onset time and sleep offset time in $\Delta t$, which were denoted as *Std_dur*, *Std_onset*, and *Std_offset*, respectively.

### Sleep Quality

In this paper, we used features of sleep efficiency, awakenings, and weekend catch-up sleep to describe sleep quality.

*Sleep efficiency*. The definition of sleep efficiency is the ratio of sleep duration to time in bed [31]. The mean sleep efficiency in $\Delta t$ was denoted as *Efficiency*.

*Awakenings* (> 5 min). The definition of awakenings (> 5 min) for one night is the number of episodes in which an individual is awake for more than 5 minutes [31]. The average number of awakenings in $\Delta t$ was denoted as *Awake_5*.

*Weekend catch-up sleep*. Weekend catch-up sleep duration is an indicator of insufficient weekday sleep, which might be associated with depression level [35]. A longer sleep duration during the weekend compared with weekdays may reflect the actual sleep needed [36]. Therefore, we calculated the mean sleep duration difference between weekend and weekdays in $\Delta t$, which was denoted as *WKD_diff*.

### Insomnia

A review of several longitudinal studies suggested that insomnia is bidirectionally related to depression [6]. According to the diagnostic features provided in [37], insomnia manifests as initial insomnia (difficulty initiating sleep at bedtime), middle insomnia (frequent or prolonged awakening throughout the night), and late insomnia (early-morning awakening with an inability to return to sleep).

In order to detect initial insomnia, we calculated the mean sleep onset time (first "Non-Awake" stage) in $\Delta t$ denoted as *Av_onset*. Since initial insomnia can lead to late sleep onset time, we calculated the percentage of days with late sleep onset time in $\Delta t$. We tried three thresholds of sleep onset later than 12 pm, 1 am, and 2 am, which were denoted as *After_12*, *After_1*, and *After_2*, respectively. We also designed a feature to count the number of days in which participants may experience middle insomnia. We define middle insomnia to be whether the sleep duration is less than 6 hours, and there is a prolonged awakening ($\geq$ 30 mins) during the night. The percentage of days with potential middle insomnia in the feature window was denoted as *M_Insomnia*. For late insomnia, the mean sleep offset time in $\Delta t$ was calculated, which was denoted as *Av_offset*.

### Hypersomnia

Hypersomnia can be another symptom of depression [7]. The hypersomnia criteria used in [38] is "sleep more than 10h/day, three days a week". In this paper, the percentage of days with sleep duration greater than 10 hours was extracted in $\Delta t$, which was denoted as *Dur_10*.

**Table 1. A list of sleep features used in this study and their short description.**

|  | Feature | Description |
|---|---|---|
| **Sleep Architecture** |  |  |
|  | Sleep_dur | Mean sleep duration |
|  | Time_bed | Mean time in bed |
|  | Deep_pct | Mean deep sleep proportion |
|  | Light_pct | Mean light sleep proportion |
|  | REM_pct | Mean REM sleep proportion |
|  | NREM_pct | Mean NREM sleep proportion |
|  | Awake_pct | Mean awake proportion |
|  | REM_L | Mean REM latency time |
| **Sleep Stability** |  |  |
|  | Std_dur | Standard deviation of sleep duration |
|  | Std_onset | Standard deviation of sleep onset time |
|  | Std_offset | Standard deviation of sleep offset time |
| **Sleep Quality** |  |  |
|  | Efficiency | Mean sleep efficiency |
|  | Awake_5 | Mean number of awakenings (> 5 mins) per night |
|  | WKD_diff | Sleep duration difference between weekend and weekdays |
| **Insomnia** |  |  |
|  | Av_onset | Mean sleep onset time |
|  | After_12 | Percentage of days with sleep onset time later than 12 pm |
|  | After_1 | Percentage of days with sleep onset time later than 1 am |
|  | After_2 | Percentage of days with sleep onset time later than 2 am |
|  | M_insomnia | Percentage of days with middle insomnia |
|  | Av_offset | Mean sleep offset time |
| **Hypersomnia** |  |  |
|  | Dur_10 | Percentage of days with sleep duration greater than 10 hours |

## Statistical method

### Data inclusion criteria

Sleep and PHQ-8 records were missing in our data cohort for a variety of expected reasons, including the participants not wearing the Fitbit wristband when they slept, participants forgetting to complete the PHQ-8, and the Fitbit wristband being damaged during the follow-up. We, therefore, specified the following inclusion criteria: (1) the PHQ-8 record should be completed, i.e., the participant answered all eight questions in the questionnaire. (2) the number of days with sleep records in the feature window should be at least 12 days (~85% of the feature window size) [39]. (3) the number of PHQ-8 records for each participant should be greater than or equal to three [40].

*Statistical Analyses*

In our study, each participant had multiple PHQ-8 records and repeated sleep measures; for this reason, we used linear mixed models which allow to accounting for both within and between-individual variability over time [41]. For each sleep feature, a three-levelled linear mixed model with a participant-specific random intercept and a site-specific random intercept was built on the entire dataset to explore the association between this sleep feature and depressive symptom severity (PHQ-8) by bivariate analysis. Then, we used two-levelled linear mixed models with participant-specific random intercepts to test these associations on three subsets (KCL, CIBER, and VUmc), separately. We also analysed the associations between sleep features and the sleep subscore, similarly. All models were adjusted for baseline age, gender, education level and annual income which were specified as fixed effects. Model assumptions were checked by the histograms of residuals and QQ plots. If the residuals are not normally distributed, the Box-Cox transformation was performed [42]. The z-test was used to evaluate the statistical significance of the coefficient of each model. All *P*-values of these tests were corrected by using the Benjamini-Hochberg method [43] for multiple comparisons, and the significance level of the corrected *P*-value was set to .05. All linear mixed models were implemented by using the "lme4" package for the R software version 3.6.1.

In order to identify and compare the relationship between self-reported sleep and self-reported depression among different study sites, Spearman correlations were calculated between the PHQ-8 score and the sleep subscore on the three study sites, separately.

An example of such a three-levelled linear mixed model is:

$$Sleep_{ijk} = \delta_{000} + V_{00k} + U_{0jk} + \beta_1(PHQ8_{ijk}) + \beta_2(age_{jk}) + \beta_3(gender_{jk}) + \beta_4(education_{jk}) + \beta_5(income_{jk}) + \varepsilon_{ijk}$$

,

where $PHQ8_{ijk}$ is the $i^{th}$ PHQ-8 score of the participant j of the site k, $Sleep_{ijk}$ is one sleep feature extracted in Δt before the $i^{th}$ PHQ-8 record of the participant j of the site k, $age_{jk}$, $gender_{jk}$, $education_{jk}$ and $income_{jk}$ are potential confounding variables of the participant j of the site k, $\varepsilon_{ijk}$ is the residual, $\delta_{000}$ is the fixed effect on intercept, $U_{0jk}$ is the random intercept of the participant j in the site k, and $V_{00k}$ is the random intercept of the site k.

## Results

### Data summary

According to our data inclusion criteria, from June 2018 to June 2020, 2812 PHQ-8 records from 368 participants collected from three study sites were included for our analysis. A summary of the sociodemographic characteristics of these participants at baseline and scores of all PHQ-8 records is shown in Table 2. The Kruskal-Wallis test was used to determine whether there were any significant differences for these characteristics between three sites. These tests revealed that, except for gender, sociodemographic characteristics and distribution of PHQ-8 scores differed between three study sites. The histograms of PHQ-8 scores of three study sites and the entire dataset are shown in Figure 1. We can observe that the KCL site had the most PHQ-8 records among these three sites. PHQ-8 scores in the CIBER site were relatively high, probably because participants in the CIBER site came from a clinical population. Figure 2 presents the Spearman correlation coefficients between all 21 sleep features. Table 3 shows the results of Spearman correlation analysis; we can observe there was a strong positive correlation between the sleep subscore and the PHQ-8 score (r=0.73, z=54.48, $P < .001$) on the entire dataset, but this correlation was relatively weaker on the VUmc data (r=0.64, z = 18.75, $P < .001$).

**Table 2. A summary of sociodemographic characteristics and PHQ-8 records of participants from the three study sites, and results of Kruskal-Wallis test on these characteristics.**

|  | KCL | CIBER | VUmc | Difference[a] |
|---|---|---|---|---|
| Participants, n | 189 | 96 | 83 | - |
| The PHQ-8 records, n | 1547 | 708 | 557 | - |
| PHQ-8 scores, median (Q1, Q3) | 8 (4, 12) | 14 (8, 19) | 9 (5, 13) | <.001 |
| The PHQ-8 score≥10, n (%) | 599(38.72%) | 492(69.49%) | 248(44.52%) | <.001 |
| Age at baseline, median (Q1, Q3) | 46 (30.25,59) | 55 (49.25,60.75) | 42 (28, 57) | <.001 |
| Female sex, n (%) | 144(76.19%) | 69(71.88%) | 65(81.93%) | .62 |
| Education,[b] n (%) |  |  |  | <.001 |
|   Degree or above | 116(61.38%) | 21(21.88%) | 40(48.19%) |  |
|   Below degree | 73(38.62%) | 75(78.13%) | 43(51.81%) |  |
| Annual income,[c] n (%) |  |  |  | .009 |
|   Less than 15000£ | 40(21.16%) | 28(29.17%) | 24(28.92%) |  |
|   15000£-40000£ | 80(42.33%) | 53(55.21%) | 34(40.96%) |  |
|   More than 40000£ | 67(35.45%) | 15(15.62%) | 14(16.87%) |  |
|   Not mentioned | 2(1.06%) | 0(0%) | 11(13.25%) |  |

[a] P value of Kruskal-Wallis test.
[b] The education levels of Spain and the Netherlands had been transformed into equivalent British education levels.
[c] The annual income levels of Spain and the Netherlands had been transformed into equivalent British levels.

**Figure 1. Histograms of the PHQ-8 scores of the three study sites and the entire dataset.**

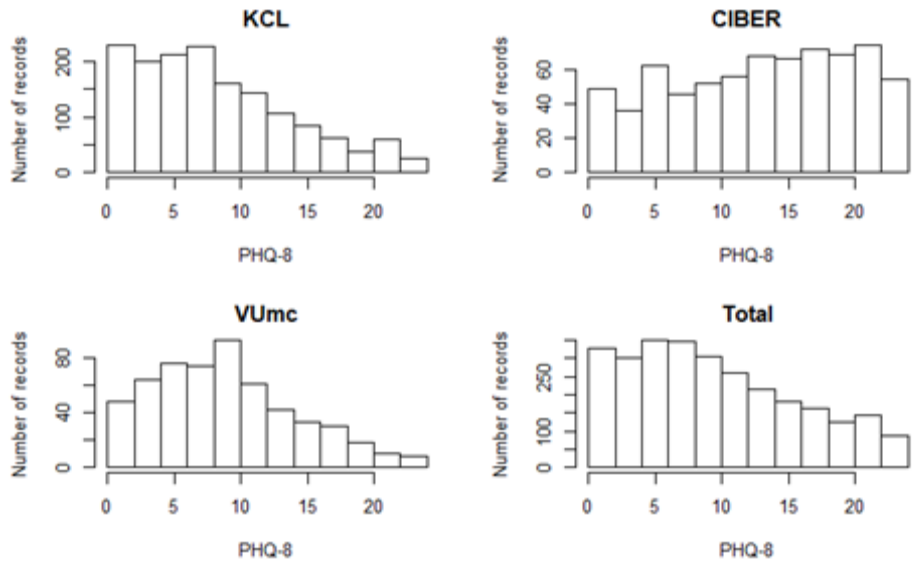

**Figure 2. A correlation plot of pairwise Spearman correlations between all sleep features. Descriptions of abbreviations of sleep features are shown in Table 1.**

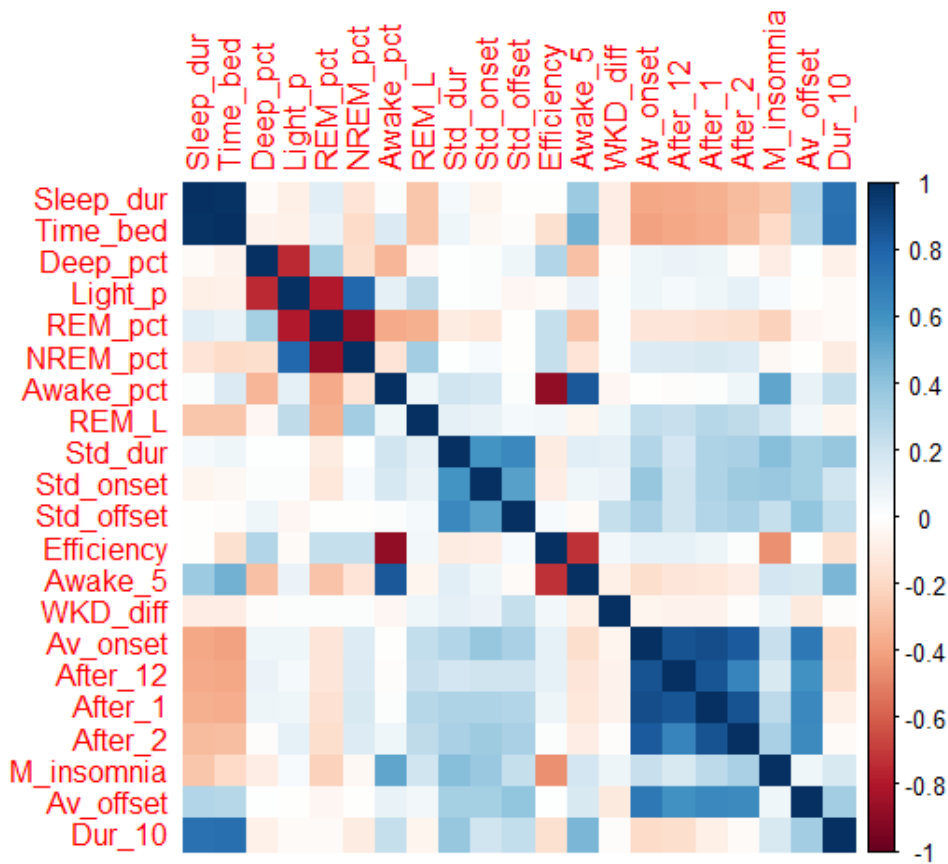

**Table 3. Spearman correlation coefficients between the PHQ-8 score and the sleep subscore[a] on the three study sites, and their 95% confidence intervals, z-test statistics, and P values.**

| | r | 95% CI[b] | z[c] | P |
|---|---|---|---|---|
| KCL | 0.74 | [0.71,0.76] | 41.99 | <.001 |
| CIBER | 0.78 | [0.75,0.81] | 32.09 | <.001 |
| VUmc | 0.64 | [0.58,0.69] | 18.75 | <.001 |
| Total | 0.73 | [0.71,0.74] | 54.48 | <.001 |

[a] The sleep subscore represents the score of sub-item 3 in the PHQ-8.
[b] Confidence interval.
[c] z-test statistics.

### Three-levelled linear mixed models on the entire dataset

Table 4 shows the results from three-levelled linear mixed regression models which reflect the associations between sleep features and the PHQ-8 score and the sleep subscore, respectively. 16 sleep features were found to be significantly correlated with the PHQ-8 score, among them, awake proportion (z = 5.45, $P < .001$), awakening times (z = 5.53, $P < .001$), insomnia (z = 4.55, $P < .001$), mean sleep offset time (z = 6.19, $P < .001$) and hypersomnia (z = 5.30, $P < .001$) were the top 5 features ranked by z-test statistics. The proportion of light sleep (*Light_pct*) and NREM sleep (*NREM_pct*), and sleep efficiency (*Efficiency*) were significantly and negatively correlated with the PHQ-8 score, whereas the rest of significant features were positively correlated with the PHQ-8 score.

For sleep subscore, we can notice that deep sleep proportion (*Deep_pct*), REM sleep proportion (*REM_pct*) and sleep efficiency (*Efficiency*) were significantly and negatively correlated with the sleep subscore, whereas features of the proportion of awake (*Awake_pct*), unstable sleep (*Std_dur*, *Std_onset*, *Std_offset*), awakening times (*Awake_5*), weekend catch-up sleep (*WKD_diff*), late sleep onset time (*Av_onset*, *After_12*, *After_1*, *After_2*), sleep offset time (*Av_offset*), insomnia (*M_insomnia*), and hypersomnia (*Dur_10*) were significantly and positively correlated with the sleep subscore.

**Table 4. Slope coefficient estimates, 95% confidence intervals, z-test statistics, and *P* values from three-levelled linear mixed models on the entire dataset for exploring associations between sleep features[a] and the PHQ-8 score and the sleep subscore[b].**

| | The PHQ-8 score | | | | The sleep subscore | | | |
|---|---|---|---|---|---|---|---|---|
| | Coeff.[c] | 95% CI[d] | z[e] | *P* | Coeff. | 95% CI | z | *P* |
| Sleep_dur | 0.013 | [0.006, 0.019] | 3.93 | <.001 | -0.004 | [-0.034, 0.025] | -0.28 | .78 |
| Time_bed | 0.016 | [0.009, 0.023] | 4.45 | <.001 | 0.005 | [-0.028, 0.038] | 0.29 | .77 |
| Deep_pct | -0.007 | [-0.026, 0.011] | -0.75 | .45 | -0.104 | [-0.191, -0.017] | -2.34 | .02 |
| Light_pct | -0.032 | [-0.064, -0.001] | -2.02 | .04 | 0.090 | [-0.057, 0.237] | 1.20 | .23 |
| REM_pct | 0.003 | [-0.021, 0.027] | 0.25 | .80 | -0.125 | [-0.238, -0.012] | -2.17 | .03 |
| NREM_pct | -0.038 | [-0.062, -0.014] | -3.12 | .002 | -0.014 | [-0.127, 0.098] | -0.25 | .80 |
| Awake_pct | 0.035 | [0.022, 0.048] | 5.45 | <.001 | 0.139 | [0.079, 0.199] | 4.58 | <.001 |
| REM_L | 0.034 | [-0.021, 0.088] | 1.21 | .23 | 0.085 | [-0.178, 0.347] | 0.63 | .53 |
| Std_dur | 0.008 | [0.004, 0.012] | 4.07 | <.001 | 0.047 | [0.028, 0.067] | 4.77 | <.001 |
| Std_onset | 0.012 | [0.004, 0.019] | 3.11 | .002 | 0.060 | [0.022, 0.097] | 3.13 | .002 |
| Std_offset | 0.012 | [0.005, 0.018] | 3.58 | <.001 | 0.069 | [0.037, 0.100] | 4.26 | <.001 |
| Efficiency | -0.025 | [-0.037, -0.012] | -3.91 | <.001 | -0.108 | [-0.167, -0.050] | -3.65 | <.001 |
| Awake_5 | 0.016 | [0.010, 0.022] | 5.53 | <.001 | 0.038 | [0.011, 0.065] | 2.77 | .006 |
| WKD_diff | 0.134 | [0.039, 0.230] | 2.76 | .006 | 0.747 | [0.255, 1.240] | 2.98 | .003 |
| Av_onset | 0.007 | [-0.001, 0.015] | 1.71 | .09 | 0.078 | [0.040, 0.115] | 4.03 | <.001 |
| After_12 | 0.101 | [-0.078, 0.280] | 1.10 | .27 | 1.600 | [0.766, 2.435] | 3.76 | <.001 |
| After_1 | 0.267 | [0.099, 0.434] | 3.12 | .002 | 2.062 | [1.281, 2.843] | 5.18 | <.001 |
| After_2 | 0.253 | [0.117, 0.389] | 3.65 | <.001 | 1.690 | [1.054, 2.326] | 5.21 | <.001 |
| M_insomnia | 0.370 | [0.211, 0.530] | 4.55 | <.001 | 2.373 | [1.595, 3.151] | 5.98 | <.001 |
| Av_offset | 0.025 | [0.017, 0.033] | 6.19 | <.001 | 0.097 | [0.060, 0.135] | 5.10 | <.001 |
| Dur_10 | 0.309 | [0.195, 0.423] | 5.30 | <.001 | 0.909 | [0.357, 1.462] | 3.23 | .001 |

[a] The definitions of sleep features in this table are shown in Table 1.
[b] The sleep subscore represents the score of sub-item 3 in the PHQ-8.
[c] Slope coefficient estimates for all sleep features
[d] Confidence interval
[e] z is z-test statistics

### Two-levelled linear mixed models on different research sites

Table 5 provides the results from two-levelled linear mixed models which show the associations between sleep features and the PHQ-8 score on different research sites separately. On the KCL data, most associations between sleep features and depression were consistent with the results on the entire dataset. On the CIBER data, some features were no longer significantly correlated with the PHQ-8 score, especially features related to sleep onset time (*Av_onset*, *Std_onset*, *After_12*, *After_1*, *After_2*). However, on the VUmc data, most features lost their significance except features of sleep duration (*Sleep_dur*), time in bed (*Time_bed*), REM latency (*REM_L*), and awakenings (*Awake_5*).

Table 6 shows associations between sleep features and the sleep subscore on different research sites. The significance of correlations between sleep features and the sleep subscore were different among the three study sites. Notably, insomnia feature (*M_insomnia*) and at least one feature of sleep stability were significantly positively correlated with sleep subscore on the data of all three sites.

**Table 5. Coefficient estimates, 95% confidence intervals, and *P* values from two-levelled linear mixed models on the three study sites for exploring associations between sleep features[a] and the PHQ-8 score.**

| | KCL | | | CIBER | | | VUmc | | |
|---|---|---|---|---|---|---|---|---|---|
| | Coeff.[b] | 95% CI[c] | *P* | Coeff. | 95% CI | *P* | Coeff. | 95% CI | *P* |
| Sleep_dur | 0.013 | [0.005, 0.020] | .001 | 0.016 | [-0.001, 0.033] | .06 | 0.011 | [0.000, 0.022] | .049 |
| Time_bed | 0.016 | [0.008, 0.024] | <.001 | 0.021 | [0.002, 0.040] | .03 | 0.013 | [0.001, 0.025] | .04 |
| Deep_pct | -0.005 | [-0.028, 0.018] | .69 | 0.024 | [-0.022, 0.071] | .31 | -0.037 | [-0.074, 0.001] | .06 |
| Light_pct | -0.046 | [-0.087, -0.006] | .03 | -0.081 | [-0.155, -0.007] | .03 | 0.019 | [-0.043, 0.082] | .55 |
| REM_pct | 0.013 | [-0.018, 0.043] | .43 | 0.015 | [-0.042, 0.071] | .62 | -0.007 | [-0.055, 0.041] | .77 |

| | | | | | | | | |
|---|---|---|---|---|---|---|---|---|
| NREM_pct | -0.049 | [-0.080, -0.018] | .002 | -0.060 | [-0.116, -0.005] | .04 | -0.016 | [-0.062, 0.030] | .50 |
| Awake_pct | 0.037 | [0.020, 0.054] | <.001 | 0.043 | [0.015, 0.071] | .003 | 0.022 | [-0.003, 0.047] | .09 |
| REM_L | 0.019 | [-0.049, 0.088] | .58 | 0.106 | [-0.026, 0.237] | .12 | -0.126 | [-0.231, -0.020] | .02 |
| Std_dur | 0.008 | [0.003, 0.013] | .001 | 0.009 | [0.000, 0.019] | .06 | 0.002 | [-0.006, 0.010] | .62 |
| Std_onset | 0.007 | [-0.002, 0.016] | .14 | 0.019 | [-0.001, 0.039] | .06 | 0.001 | [-0.011, 0.013] | .93 |
| Std_offset | 0.009 | [0.001, 0.017] | .03 | 0.019 | [0.002, 0.036] | .03 | 0.003 | [-0.008, 0.015] | .56 |
| Efficiency | -0.025 | [-0.041, -0.008] | .004 | -0.043 | [-0.071, -0.016] | .002 | -0.012 | [-0.037, 0.013] | .34 |
| Awake_5 | 0.014 | [0.006, 0.022] | <.001 | 0.022 | [0.009, 0.035] | .001 | 0.016 | [0.005, 0.027] | .01 |
| WKD_diff | 0.211 | [0.084, 0.339] | .001 | 0.071 | [-0.126, 0.268] | .48 | 0.077 | [-0.144, 0.299] | .49 |
| Av_onset | 0.010 | [0.000, 0.020] | .047 | 0.004 | [-0.018, 0.025] | .74 | -0.005 | [-0.021, 0.010] | .52 |
| After_12 | 0.167 | [-0.063, 0.397] | .16 | 0.098 | [-0.323, 0.520] | .65 | -0.216 | [-0.595, 0.162] | .26 |
| After_1 | 0.334 | [0.123, 0.545] | .002 | 0.123 | [-0.305, 0.552] | .57 | 0.089 | [-0.235, 0.413] | .59 |
| After_2 | 0.328 | [0.150, 0.506] | <.001 | 0.078 | [-0.260, 0.417] | .65 | 0.123 | [-0.118, 0.364] | .32 |
| M_insomnia | 0.472 | [0.259, 0.685] | <.001 | 0.381 | [0.028, 0.734] | .04 | -0.048 | [-0.385, 0.289] | .78 |
| Av_offset | 0.029 | [0.018, 0.039] | <.001 | 0.024 | [0.004, 0.043] | .02 | 0.012 | [-0.004, 0.029] | .14 |
| Dur_10 | 0.331 | [0.191, 0.472] | <.001 | 0.340 | [0.052, 0.627] | .02 | 0.181 | [-0.051, 0.413] | .13 |

[a] The definitions of sleep features in this table are shown in Table 1.
[b] Slope coefficient estimates for all sleep features
[c] Confidence interval

**Table 6. Coefficient estimates, 95% confidence intervals, and *P* values from two-levelled linear mixed models on the three study sites for exploring associations between sleep features[a] and the sleep subscore[b].**

| | KCL | | | CIBER | | | VUmc | | |
|---|---|---|---|---|---|---|---|---|---|
| | Coeff.[c] | 95% CI[d] | P | Coeff. | 95% CI | P | Coeff. | 95% CI | P |
| Sleep_dur | 0.015 | [-0.021, 0.050] | .41 | -0.035 | [-0.116, 0.047] | .41 | -0.017 | [-0.070, 0.035] | .52 |
| Time_bed | 0.026 | [-0.013, 0.066] | .19 | -0.025 | [-0.116, 0.065] | .58 | -0.015 | [-0.074, 0.043] | .61 |
| Deep_pct | -0.027 | [-0.134, 0.081] | .63 | -0.196 | [-0.412, 0.020] | .07 | -0.191 | [-0.369, -0.014] | .04 |
| Light_pct | -0.024 | [-0.213, 0.166] | .81 | 0.098 | [-0.250, 0.445] | .58 | 0.312 | [0.016, 0.608] | .04 |
| REM_pct | -0.116 | [-0.260, 0.028] | .12 | -0.037 | [-0.304, 0.230] | .79 | -0.169 | [-0.398, 0.060] | .15 |
| NREM_pct | -0.048 | [-0.194, 0.098] | .52 | -0.123 | [-0.389, 0.143] | .37 | 0.125 | [-0.096, 0.346] | .27 |
| Awake_pct | 0.165 | [0.085, 0.245] | <.001 | 0.150 | [0.020, 0.280] | .02 | 0.049 | [-0.073, 0.170] | .43 |
| REM_L | 0.073 | [-0.255, 0.401] | .66 | 0.146 | [-0.494, 0.787] | .65 | -0.171 | [-0.683, 0.340] | .51 |
| Std_dur | 0.046 | [0.022, 0.071] | <.001 | 0.046 | [-0.002, 0.094] | .06 | 0.043 | [0.004, 0.082] | .03 |
| Std_onset | 0.028 | [-0.015, 0.070] | .21 | 0.089 | [-0.018, 0.195] | .10 | 0.079 | [0.020, 0.139] | .01 |
| Std_offset | 0.046 | [0.008, 0.084] | .02 | 0.109 | [0.022, 0.195] | .01 | 0.072 | [0.016, 0.127] | .01 |
| Efficiency | -0.118 | [-0.196, -0.041] | .003 | -0.152 | [-0.280, -0.024] | .02 | -0.044 | [-0.162, 0.074] | .46 |
| Awake_5 | 0.047 | [0.011, 0.083] | .01 | 0.037 | [-0.022, 0.097] | .22 | 0.013 | [-0.042, 0.067] | .65 |
| WKD_diff | 1.169 | [0.534, 1.804] | <.001 | 0.210 | [-0.864, 1.284] | .70 | 0.283 | [-0.830, 1.395] | .62 |
| Av_onset | 0.055 | [0.008, 0.101] | .02 | 0.075 | [-0.023, 0.172] | .13 | 0.128 | [0.054, 0.202] | .001 |
| After_12 | 0.945 | [-0.125, 2.015] | .08 | 1.716 | [-0.197, 3.628] | .08 | 2.984 | [1.187, 4.782] | .001 |
| After_1 | 1.837 | [0.855, 2.820] | <.001 | 1.491 | [-0.459, 3.441] | .14 | 2.981 | [1.452, 4.509] | <.001 |
| After_2 | 1.723 | [0.886, 2.560] | <.001 | 0.667 | [-0.863, 2.197] | .39 | 2.527 | [1.396, 3.657] | <.001 |
| M_insomnia | 2.302 | [1.274, 3.329] | <.001 | 2.777 | [1.070, 4.485] | .001 | 1.823 | [0.180, 3.465] | .03 |
| Av_offset | 0.102 | [0.053, 0.150] | <.001 | 0.048 | [-0.040, 0.135] | .29 | 0.133 | [0.056, 0.210] | .001 |
| Dur_10 | 1.057 | [0.387, 1.728] | .002 | 0.576 | [-0.844, 1.995] | .43 | 0.706 | [-0.411, 1.823] | .22 |

[a] The definitions of sleep features in this table are shown in Table 1.

[b] The sleep subscore represents the score of sub-item 3 in the PHQ-8.

[c] Slope coefficient estimates for all sleep features.

[d] Confidence interval.

## Discussion

### Principal Findings

In this study, we extracted 21 sleep features through Fitbit data to quantitatively describe the participant's sleep characteristics in five categories (*sleep architecture*, *sleep stability*, *sleep quality*, *insomnia*, and *hypersomnia*) that were associated with the severity of depression. Along with the depressive status worsening, the following changes may be seen in the past 2 weeks: (1) the proportion of light/NREM sleep decreased, and the proportion of wakefulness during sleep increased (*sleep architecture*); (2) sleep duration/onset/offset were unstable (*sleep stability*); (3) reduced sleep efficiency, more awakenings during sleep, and longer weekend sleep catch-up sleep (*sleep quality*); (4) more days with insomnia (*insomnia*); (5) more days with hypersomnia (*hypersomnia*). From Table 4, it illustrated that our sleep features of these five categories could both reflect the participants' sleep condition (the sleep subscore) and depressive symptom severity (the PHQ-8 score) of past two weeks.

### Potential factors affecting associations

We evaluated our models on these three research sites separately. From Table 5 and Table 6, we can notice that the associations between sleep features and the PHQ-8 score/sleep subscore varied across different sites. There were several factors which may affect the associations. First, the populations of the three sites were significantly different (Table 2). For example, participants in the CIBER site came from a clinical population, and their average age was oldest, so one speculation is that there was less difference between their weekday sleep and weekend sleep for inpatients or people in retirement. Therefore, this may be the reason why the feature of weekend catch-up sleep (*WKD_diff*) lost significance on the CIBER data. In addition, the reduced significance of features related to sleep onset and offset time on CIBER site might be related to the regular sleep pattern in CIBER site favours going to bed later, as seen in our previous study [44].

The associations between of sleep features and the sleep subscore on the VUmc data (Table 6) were similar to that on the entire dataset (Table 4), which demonstrated our sleep features have the same ability to capture the sleep condition of participants on the VUmc data. However, the significance of associations between these sleep features and the PHQ-8 score was reduced on the VUmc data (Table 6). One possible reason is that, from Table 3, the correlation between the sleep subscore and the PHQ-8 score on the VUmc data (r=0.64) was weaker than other two study sites (KCL: r=0.74 and CIBER: r=0.78) which may be caused by confounding variables that we did not consider or record in the VUmc population, such as medication and occupational status.

The sample size and the heterogeneity of the dataset were other possible factors which may affect results. It can be noticed from Table 2, the KCL site had the most PHQ-8 records, whereas VUmc had the least data. As depression manifests itself in distinctive symptoms on different people, it may be difficult to fully explore the associations between sleep and depression on a relatively smaller dataset (VUmc). For example, hypersomnia is specifically related to bipolar patients [7, 8]; therefore, if the dataset did not contain enough bipolar patients or bipolar patients were not in depressive episodes when they completed their PHQ-8 records, it was hard to find the association between hypersomnia and depression.

**Figure 3. The PHQ-8 scores and a select 4 sleep features of one participant with an obvious increasing trend in PHQ-8 score at 13$^{th}$ PHQ-8 record. Descriptions of abbreviations of sleep features in this figure are shown in Table 1.**

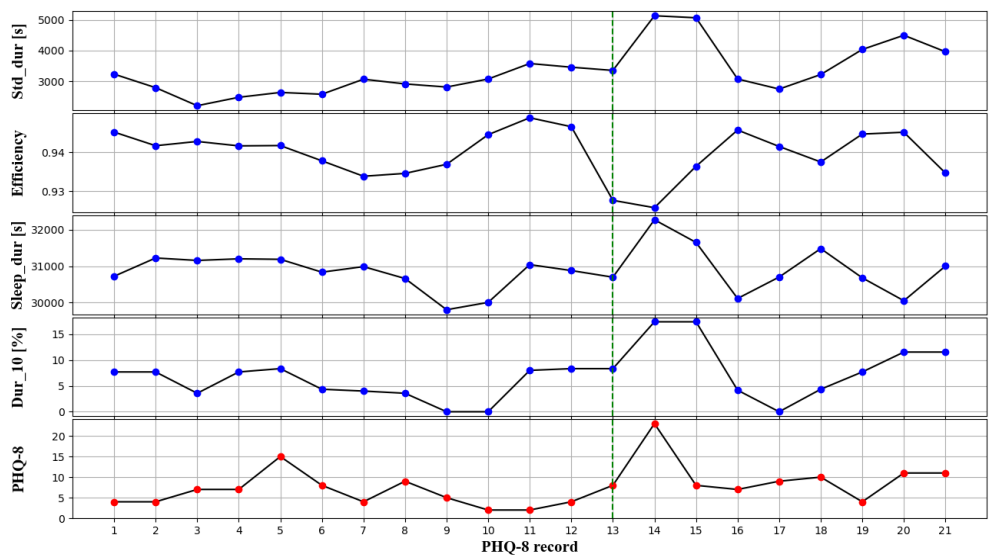

**Comparison with Prior Work**

Our study has a relatively larger sample size and a longer follow-up duration than previous studies on monitoring depression by using wearable devices and mobile phones [19-21]. Each participant has multiple PHQ-8 records and repeated measurements of sleep, so we can not only explore the relationships between sleep and depression between-individuals but also find the associations within-individuals by using the linear mixed model. Figure 3 is an example of a possible depression relapse of one participant. It showed an obvious increasing trend in PHQ-8 scores at the 13$^{th}$ PHQ-8 record of this participant. We can observe the sleep features in Figure 3 are significantly correlated with PHQ-8 score, especially the feature of hypersomnia (*Dur_10*) which indicates this participant may have hypersomnia during the period of depression relapse. This indicates that the sleep features extracted in this paper have the potential to be the biomarkers of depression.

We also compared our findings with previous studies which used other measurements to assess sleep, such as PSG and sleep questionnaires. Although the sample size, population, measurements, duration of these studies are different, the comparison may help to find more general associations between sleep and depression. Table 7 provides a summary of the comparison. Several longitudinal studies based on sleep questionnaires have shown that insomnia and hypersomnia are both symptoms of depression [6, 45], which is the same as the results of our paper. S.-G. Kang et al. found the weekend catch-up sleep duration was significantly positively correlated with the severity of depression by analysing the self-sleep questionnaires of 4553 Korean adolescents [35], which is consistent with the finding in our paper. A sleep report has shown that higher sleep efficiency, more deep sleep and fewer awakenings after sleep onset represents better sleep quality [31], which is also consistent with the relationship we found between deep sleep proportion, awake proportion, and awakenings (>5 mins) with sleep subscore. A review showed that according to PSG research, the shortened REM latency and the increased proportion of REM sleep are biological marks of depression relapse [9]. However, relationships between depressive symptom severity with the REM sleep proportion and REM latency were not significant in our results.

**Table 7. A summary of the comparisons with previous studies using other measurements to assess sleep.**

| Type of feature | Findings in previous studies | Consistent[a] | Measurement | Ref. |
|---|---|---|---|---|
| Insomnia | Insomnia is bidirectionally related to depression | Yes | Questionnaire | [6] |
| Hypersomnia | Prevalence of hypersomnia is high in depressed patients | Yes | Questionnaire | [45] |
| Weekend catch-up sleep | Weekend catch-up sleep duration is significantly positively correlated with the severity of depression | Yes | Questionnaire | [35] |
| Deep sleep proportion | More deep sleep represents higher sleep quality | Yes | Questionnaire | [31] |
| Awakening proportion, Awakenings (>5 mins) | Fewer awakenings after sleep onset represents better sleep quality | Yes | Questionnaire | [31] |
| Sleep efficiency | Higher sleep efficiency represents better sleep quality | Yes | Questionnaire | [31] |
| REM sleep proportion | Increased REM sleep duration can be biomarkers of depression | No | PSG | [9] |
| REM latency | Shortened REM latency can be biomarkers of depression | No | PSG | [9] |

[a] Whether it is consistent with our findings

### Limitations

Missing data is the major hindrance to our study. For various reasons, there were many missing records of sleep. We set the completion rate of sleep records greater than 85% (12 days) as one of the data inclusion criteria. However, the optimum threshold is unclear, which needs to be further studied in future research. Besides, missingness could also be associated with the depressive status and be a useful marker of relapse of depression; for example, participants may not feel like complying if they are feeling depressed. In future research, we will consider missingness as a potential feature.

Although we adjusted our models for age, gender, education level, and annual income, it is hard to consider all potential confounding variables. For example, some participants with sleep disorders may take sleep medications. Sleep medications have a significant influence on the features of sleep. Unfortunately, there was no daily record of whether the participant took medication. This confounding variable may affect the result.

The data of sleep stages used in this paper were provided by Fitbit wristband. According to their validation studies, the Fitbit wristband showed promise in detecting sleep-wake states, but with limitations in other sleep stages estimation [26-28]. Probably, for this reason, the features of REM proportion and REM latency in our paper did not show significant relationships with depressive symptoms. Although there are some limitations of Fitbit data, it provides a means to investigate sleep characteristic in home settings.

In feature extraction, we did not consider the impact of individual circumstances on sleep features. For example, some participants may need to shift work at night, which our features are unable to capture. We will consider the impact of sleep habits and lifestyles on sleep features in the future.

In this paper, we only performed bivariate analysis, i.e. separately analysing the association between each feature and the PHQ-8 score. The combination of features and non-linear relationships were not considered. We will try to apply machine/deep learning models to predict the severity of depression by using our sleep features in future research.

Since all features in this paper were extracted manually, information of sleep characteristics may not be fully explored. For future work, we will try to use unsupervised learning, like autoencoders [46], to extract the latent features automatically.

## Conclusions

Although consumer wearable devices may not be a substitute for PSG to assess sleep quality accurately, we demonstrate that some derived sleep features extracted from these wearable devices show potential for remote measurement of sleep and consequently as a biomarker of depression in real-world settings. These findings may provide the basis for the development of clinical tools that could be used to passively monitor disease state and trajectory with minimal burden on the participant.

## Acknowledgments

The RADAR-CNS project has received funding from the Innovative Medicines Initiative 2 Joint Undertaking under grant agreement No 115902. This Joint Undertaking receives support from the European Union's Horizon 2020 research and innovation programme and EFPIA, www.imi.europa.eu. This communication reflects the views of the RADAR-CNS consortium and neither IMI nor the European Union and EFPIA are liable for any use that may be made of the information contained herein.

This paper represents independent research part-funded by the National Institute for Health Research (NIHR) Biomedical Research Centre at South London and Maudsley NHS

## Abbreviations

RADAR: Remote Assessment of Disease and Relapse
MDD: Major Depressive Disorder
PHQ-8: Patient Health Questionnaire 8-item
KCL: King's College London (London, UK)
CIBER: Centro de Investigación Biomédican en Red (Barcelona, Spain)
VUmc: Vrije Universiteit Medisch Centrum (Amsterdam, The Netherlands)